\documentclass[letterpaper]{ptephy_v1}
\preprintnumber{XXXX-XXXX} 

\usepackage{subfig} 
\usepackage{type1cm}

\begin{document}

\title{
Structure near $K^-$+$p$+$p$ threshold in the in-flight $^3$He$(K^-,\Lambda p)n$ reaction}


\author{\textbf{J-PARC E15 Collaboration}\\Y.~Sada} 
\affil{\small{Research Center for Nuclear Physics (RCNP), Osaka University,
 Osaka, 567-0047, Japan}  }

\author[1]{S.~Ajimura}
\author{M.~Bazzi}
\affil{ Laboratori Nazionali di Frascati dell' INFN, I-00044 Frascati, Italy }
\author{G.~Beer}
\affil{ Department of Physics and Astronomy, University of Victoria,
 Victoria BC V8W 3P6, Canada }
\author{H.~Bhang}
\affil{ Department of Physics, Seoul National University, Seoul,
 151-742, South Korea }
\author{M.~Bragadireanu},
\affil{ National Institute of Physics and Nuclear Engineering - IFIN
 HH, Romania }
\author{P.~Buehler}
\affil{ Stefan-Meyer-Institut f\"{u}r subatomare Physik, A-1090 Vienna,
 Austria }
\author[7,9]{L.~Busso}
\affil{ INFN Sezione di Torino, Torino, Italy }
\author[6]{M.~Cargnelli}
\author[4]{S.~Choi}
\author[2]{C.~Curceanu}
\author{S.~Enomoto}
\affil{ High Energy Accelerator Research Organization (KEK), Tsukuba,
 305-0801, Japan }
\author[7,9]{D.~Faso}
\affil{ Dipartimento di Fisica Generale, Universita' di Torino, Torino,
 Italy }

\author{H.~Fujioka}
\affil{ Department of Physics, Kyoto University, Kyoto, 606-8502, Japan }
\author{Y.~Fujiwara}
\affil{ Department of Physics, The University of Tokyo, Tokyo,
 113-0033, Japan }
\author{T.~Fukuda}
\affil{ Laboratory of Physics, Osaka Electro-Communication University,
 Osaka, 572-8530, Japan }
\author[2]{C.~Guaraldo}
\author{T.~Hashimoto}
\affil{ RIKEN Nishina Center, RIKEN, Wako, 351-0198, Japan }
\author[11]{R.~S.~Hayano}
\author[1]{T.~Hiraiwa}
\author[8]{M.~Iio}
\author[2]{M.~Iliescu}
\author[1]{K.~Inoue}
\author[10]{Y.~Ishiguro}
\author[11]{T.~Ishikawa}
\author[8]{S.~Ishimoto}
\author[6]{T.~Ishiwatari}
\author[13]{K.~Itahashi}

\author[8]{M.~Iwai}
\author[14,13]{M.~Iwasaki}
\affil{ Department of Physics, Tokyo Institute of Technology, Tokyo,
 152-8551, Japan }
\author[13]{Y.~Kato}
\author{S.~Kawasaki}
\affil{ Department of Physics, Osaka University, Osaka, 560-0043, Japan }
\author{P.~Kienle\thanks{deceased}}
\affil{ Technische Universit\"{a}t M\"{u}nchen, D-85748, Garching, Germany }
\author[14]{H.~Kou}
\author[13]{Y.~Ma}
\author[6]{J.~Marton}
\author{Y.~Matsuda}
\affil{ Graduate School of Arts and Sciences, The University of Tokyo,
 Tokyo, 153-8902, Japan }
\author[12]{Y.~Mizoi}
\author[7]{O.~Morra}
\author[10]{T.~Nagae}
\author[1]{H.~Noumi}
\author[13,1]{H.~Ohnishi}
\author[13]{S.~Okada}
\author[13]{H.~Outa}
\author[2]{K.~Piscicchia}
\author[2]{A.~Romero~Vidal}
\author[15]{A.~Sakaguchi}
\author[13]{F.~Sakuma}
\author[13]{M.~Sato}
\author[2]{A.~Scordo}
\author[8]{M.~Sekimoto}
\author[2]{H.~Shi}
\author[2,5]{D.~Sirghi}
\author[2,5]{F.~Sirghi}
\author[6]{K.~Suzuki}
\author[8]{S.~Suzuki}
\author[11]{T.~Suzuki}
\author{K.~Tanida}
\affil{ ASRC, Japan Atomic Energy Agency, Ibaraki 319-1195, Japan}
\author{H.~Tatsuno}
\affil{Department of Chemical Physics, Lund University, Lund, 221 00, Sweden}
\author[14]{M.~Tokuda}
\author[1]{D.~Tomono}
\author[8]{A.~Toyoda}
\author{K.~Tsukada}
\affil{ Department of Physics, Tohoku University, Sendai,
 980-8578, Japan }
\author[2,21]{O.~Vazquez~Doce}
\affil{ Excellence Cluster Universe, Technische Universit\"{a}t
M\"{u}nchen, D-85748, Garching, Germany }
\author[6]{E.~Widmann}
\author[6]{B.~K.~Wuenschek}
\author[15]{T.~Yamaga}
\author[11,13]{T.~Yamazaki}
\author{H.~Yim}
\affil{ Korea Institute of Radiological and Medical Sciences (KIRAMS),
 Seoul, 139-706, South Korea@
\email{sada@rcnp.osaka-u.ac.jp}}
\author[13]{Q.~Zhang}
\author[6]{J.~Zmeskal}






\begin{abstract}%

To search for an S= -1 di-baryonic state which decays to $\Lambda p$, 
the $ {\rm{}^3He}(K^-,\Lambda p)n_{missing}$ reaction was studied at 1.0 GeV/$c$. 
Unobserved neutrons were kinematically identified from the missing mass $M_X$ of the $ {\rm{}^3He}(K^-,\Lambda p)X$ reaction in order to have a large acceptance for the $\Lambda pn$ final state.
The observed $\Lambda p n$ events, distributed widely over the kinematically allowed region of the Dalitz plot, establish that the major component comes from a three nucleon absorption process.
A concentration of events at a specific neutron kinetic energy was observed in a region of low momentum transfer to the $\Lambda p$.
To account for the observed peak structure, the simplest S-wave pole was assumed to exist in the reaction channel, having a Breit-Wigner form in energy and with a Gaussian form-factor. 
A minimum $\chi^2$ method was applied to deduce its mass $M_X\ =$ 2355 $ ^{+ 6} _{ - 8}$ (stat.) $ \pm 12$ (syst.) MeV/$c^2$, and decay-width $\Gamma _X\ = $ 110 $ ^{+ 19} _{ - 17}$ (stat.) $ \pm 27$ (syst.) MeV/$c^2$, respectively. The form factor parameter $Q_X \sim$ 400 MeV/$c$ implies that the range of the interaction is about 0.5 fm.


\end{abstract}

\subjectindex{D01, D33}

\maketitle

\section{Introduction}

The $\bar{K}N$ interaction is known to be strongly attractive from low-energy  scattering data \cite{lowsca} and X-ray spectroscopy of kaonic atoms \cite{katom}.
By assuming that the $\Lambda $(1405) is a $K^-p$ bound state, the existence of a kaonic nuclear bound state has been predicted \cite{AY1, AY2}.
Observation of a kaonic nuclear bound state would provide definitive information on the $\bar{K}N$ interaction in the sub-threshold region, as well as the nature of $\Lambda$(1405).

Both theoretical and experimental advances have been made in the last decade. Especially, careful attention has been paid to the simplest kaonic nuclear $\bar{K}NN$ state.
Theoretically, all calculations predict the existence of a bound state. However, the predicted  $\bar{K}NN$ pole positions, depending on $\bar{K}N$ interaction models, are scattered. For the energy independent model (static calculation), the binding energy is 47 -- 95 MeV \cite{AY2}-\cite{tin6}. On the other hand, it becomes 9 -- 32
MeV \cite{tin6}-\cite{te2} for the energy dependent case. The widths are also widely scattered between 34 -- 110 MeV/$c^2$.

Experimentally, there are many reports on observed peak structure $\sim$100 MeV below the $\bar{K}NN$ production threshold. 
The FINUDA  group reported a peak structure in the back-to-back $\Lambda p$ invariant mass spectra via the  stopped kaon reaction on $^6$Li, $^7$Li and ${}^{12}$C targets \cite{finuda} with binding energy (B.E.)  115 $^{+6}_{-5}$ MeV, and a width ($\Gamma $) 67 $^{+14}_{-11}$ (stat.) $^{+2}_{-3}$ (syst.) MeV/$c^2$. 
The DISTO group observed $\bar{K}NN$ decaying to $\Lambda p$ in $pp$ collisions at 2.85 GeV, with a B.E. of 103 $\pm 3$ (stat.) $\pm $5 (syst. ) MeV, and $\Gamma $ of 118 $\pm $8 (stat.) $\pm $10  (syst.) MeV/$c^2$ \cite{disto}. 
In the pion induced reaction, d($\pi^+$,$K^+$) at 1.69 GeV/$c$, the E27 group observed ``$K^-pp$"-like structure in the $\Sigma ^0p$ decay mode at B.E. = 95 $^{+18}_{-17}$ (stat.) $^{+30}_{-21} $ (syst.) MeV, $\Gamma $ = 162  $^{+87}_{-45}$ (stat.) $^{+66}_{-78}$ (syst.) MeV/$c^2$ \cite{e27}. 
Conversely, no significant structure was observed in a SPring-8/LEPS $\gamma$ induced inclusive experiment \cite{leps} or in a proton-proton interaction at HADES/GSI \cite{hades}. 
Also, for the kaon stopped reaction, the other interpretations ($i.e. $  two nucleon absorption of kaons which have the final state $\Lambda p$ or $\Sigma ^0 p$) are widely discussed \cite{finuda2,kloe}. Thus, the evidence for kaonic nuclei remains controversial. 

In an attempt to clarify this situation, the E15 experiment on the $K^- + ^3{\rm He}$ reaction at $p_{K^-} =1$ GeV/$c$ is under way at J-PARC. The first physics data were accumulated in May 2013. 
By using an in-flight reaction and a light nuclear target, backgrounds from multi-nucleon absorption processes and hyperon decays are expected to be kinematically discriminated from the  $\bar{K}NN$ signal. 
The semi-inclusive forward neutron spectrum in the E15 experiment has a long sub-threshold tail reaching  $\sim $ 100 MeV below the $\bar{K} NN$ threshold, but no significant structure was seen in the deeply-bound region  \cite{hashimoto}.

In this paper, we present a study of the $K^- + ^3{\rm He}$ reaction at $p_{K^-} =1$ GeV/$c$ focusing on the $\Lambda p$ invariant mass spectrum, in particular on the $\Lambda pn$ final state for the May 2013 data.

\section{The Experimental apparatus}
Here, the experimental apparatus relevant to the $\Lambda p n$ study is briefly described.
The ongoing experiment is being conducted at the J-PARC 30 GeV syncrotron (MR).
In May 2013, typical beam intensity of the primary proton beam in MR was 30 $\times$ 10$^{12}$ per spill, 
where spill length was 2 seconds with a 6 second repetition cycle.
A high intensity proton beam from MR 
impinges on a gold target, and 1 GeV/$c$ $K^-$s are selected by the K1.8BR beamline spectrometer and purified  with an electrostatic separator. The typical $K^-$/$\pi$ ratio was 0.45.
At the online trigger level, $K^-$s are selected with an Aerogel Cherenkov counter.
The purity of the $K^-$ at this level, determined by time-of-flight (ToF) analysis, was 99 \%, with an intensity of 1.5 $\times$ 10$^5$ per spill.
The momentum resolution of the beamline spectrometer is (2.0 $\pm$ 0.5) $\times$ 10$^{-3}$ with an absolute precision of 2~MeV/$c$ at 1~GeV/$c$.
A more detailed description can be found in \cite{ptep}.

\begin{figure}[!hbtp]
\center\includegraphics[width=3.5in]{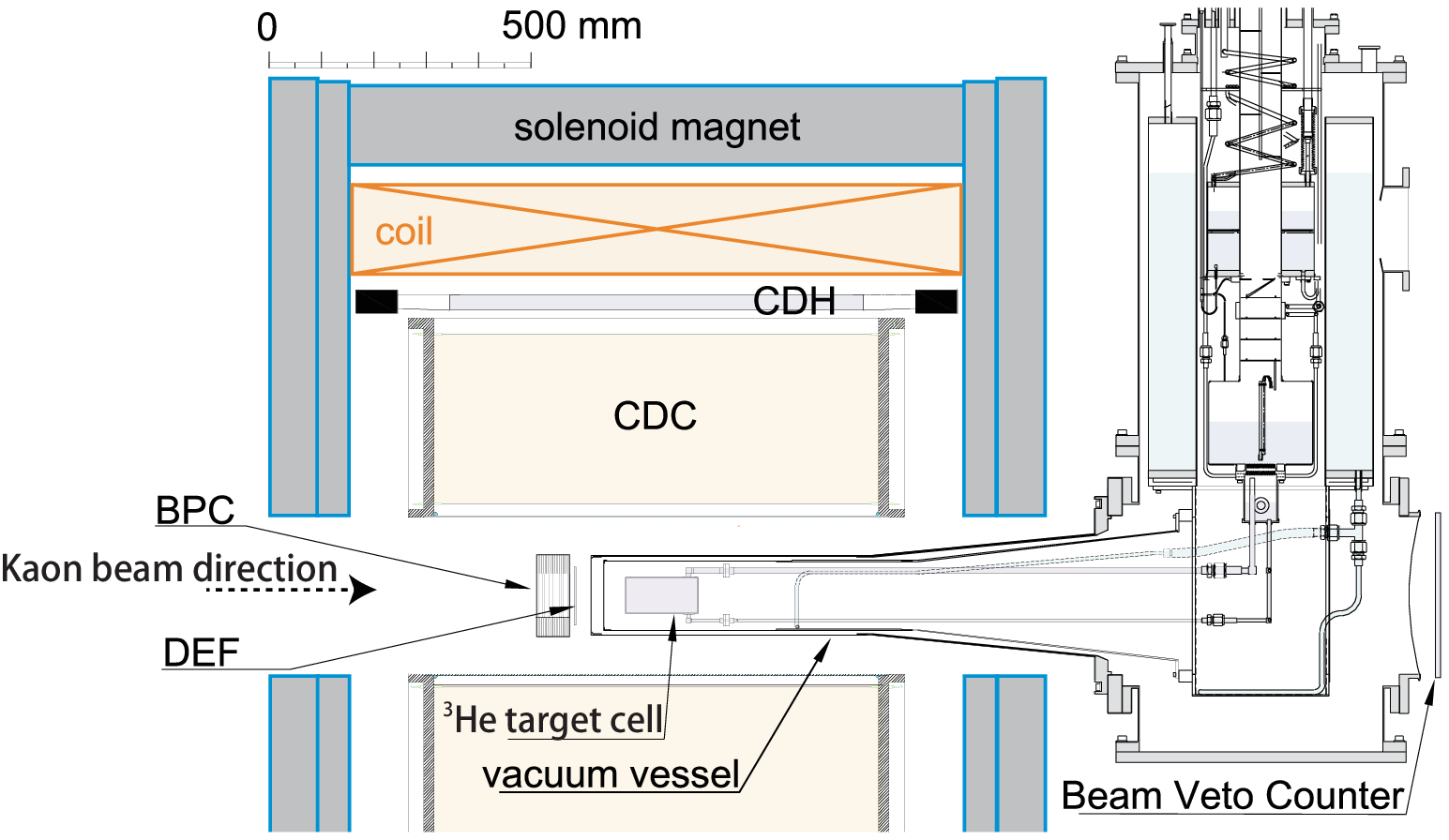}%
\caption{ Schematic diagram of detectors in the CDS and of the target system \cite{ptep}. }
\label{brdet}
\end{figure} 

 Figure \ref{brdet} shows a schematic diagram of the setup around the $^3$He target. A cylindrical target cell, 137 mm long and 68 mm in diameter, filled with liquid $^3$He, was
 placed at the final focus of the beam line. The density of the target was 0.081 g/cm$^3$ at a temperature of 1.4 K. Details of the target system are given in \cite{3hetarget}. 
To detect particles incident on the target, a small scintillator (DEF) was placed  in front of the target cell.
The reconstructed beam track from the K1.8BR beam line spectrometer was finally connected to the hit point on
a drift chamber (BPC), installed just upstream of the DEF, to improve reconstructed vertex resolution.
To measure the $K^-$ reaction products, the $^3$He target was surrounded by    
a cylindrical detector system (CDS), consisting of a Cylindrical Drift Chamber (CDC) and a Cylindrical Detector Hodoscope (CDH) operating in a uniform $\sim$ 0.7~T magnetic field.
The secondary particle tracks are reconstructed by the CDC using a helix fitting method, and their momenta are measured. Energy loss of each track, where major loss is mainly in the target region, was evaluated and corrected by the Runge-Kutta method. The transverse
momentum resolution ($\sigma_{P_t}$) is found to be 5.3 \% $P_t$ $\oplus $ 0.5 \% $/\beta$ , where $P_t$ and $\beta$ are the transverse momentum in GeV/$c$ and the velocity of the charged particle, respectively.
Particle identification (PID) was performed based on the ToF between incoming kaon timing and the CDH. Details of the CDS detectors are also described in \cite{ptep}.

The data acquisition (DAQ) trigger signal was generated by a kaon in the beamline spectrometer, a hit on the DEF counter, and two or more charged-particle hits in the CDH.  
The trigger rate was typically $\sim$ $10^3 $ per spill with a DAQ live time of $\sim$80 \%. 
During the May 2013 run period of $\sim$ 90 hours,  data corresponding to 3.4 $\times$ 10$^9$ effective kaons on the $^3$He target were accumulated and used in the following analysis.

\section{Analysis}\label{ana}
\begin{figure}[!hbtp]
\centering\includegraphics[width=3.5in]{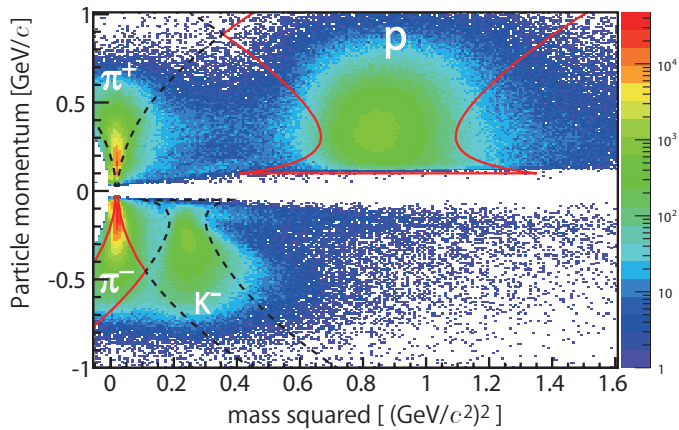}%
\caption{ PID distribution of the CDS. The cut region for each particle is defined as 2.5 $\sigma $. Overlapped regions are excluded.  }
\label{PIDwline}
\end{figure}

The $\Lambda p n$ final state was identified by establishing that the $\pi^- p$ pair came from the $\Lambda$  decay following detection of two protons and one negative pion in the CDS. The missing neutron was identified kinematically.
To reconstruct the $\Lambda p$ tracks in the CDS, precise PID was performed  based on a mass calculation using the momentum of the reconstructed track along with the CDH-based ToF information. 
Energy loss corrections based on properties of the inner CDC materials yielded improved PID functions and permitted more accurate momentum reconstructions at the reaction vertex.
The calculated mass-squared is shown as a function of momentum in Fig.~\ref{PIDwline}.
The mass-squared distribution was sliced into momentum regions and fitted with Gauss functions for each particle species. The standard deviation was used for the PID function of each particle.
The cuts defining protons, kaons and pions, shown as lines on Fig. \ref{PIDwline}, were set to 2.5 times the sigma of the respective mass-squared distributions. 
To avoid misidentification, regions of overlap were excluded, resulting in an estimated purity of 99.5 \% for protons and pions. 
 
To identify the $\pi^- p$ pair associated with a specific $\Lambda$ decay in a $\pi^- pp$ event, a Log-likelihood method was used on the product of five probability density functions: 
(1)  distance of closest approach (DCA) between a $\pi ^- p$ pair candidate for $\Lambda$ decay ($f_{DCA(\pi p)}$), (2) DCA between the kaon beam and an un-paired proton ($f_{DCA(K^- p)}$), (3) DCA between the kaon beam  and a reconstructed $\Lambda$ track ($f_{DCA(K^- \Lambda)}$), (4) DCA between the un-paired proton and the reconstructed $\Lambda$ track on the kaon beam ($f_{DCA(\Lambda p)}$),  and (5) invariant mass distribution of the $\pi ^- p $ pair candidate ($f_{M_{inv.\pi^- p}}$).
Probability density functions were calculated based on the distributions of these five quantities for a simulated $\Lambda pn$ final state. In this simulation, we assume a flat distribution (S-wave) over the three-body phase space based on the phase volume, namely:
\begin{equation}
\label{3NALpn}
\frac{d^{2}\sigma_{3\rm{NA}\left(\Lambda p n\right)}} {dT_{n}^{\,CM}d\cos\theta^{\,CM}_{n}}
\propto \rho_{3} \left( \Lambda p n \right),
\end{equation}
where $\sigma_{3\rm{NA}\left(\Lambda p n\right)}$ is 
the simulated event distribution for the $\Lambda p n$ final state, $T_{n}^{\,CM}$ is the kinetic energy of the neutron in the CM frame, and $ \cos\theta^{\,CM}_{n} $ is the neutron emission angle in the CM frame.
GEANT4 \cite{geant4} is utilized to take into account geometrical information for all detectors and their resolution for all simulations.

\begin{figure}[!hbtp]
\centering\includegraphics[width=3.5in]{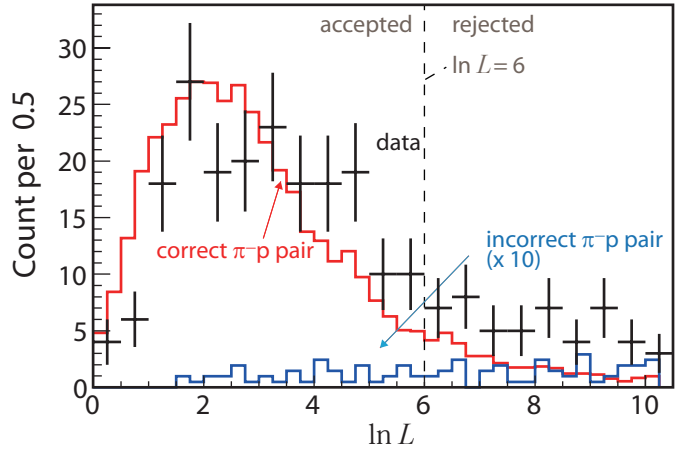}
\caption{ Likelihood function (${\rm ln}L$) distribution. Black crosses are ${\rm ln}L$ of data. Since there are two possible pairs, the pair having smaller ${\rm ln}L$ is plotted. The red histogram shows the simulated ${\rm ln}L$ of the correct $p \pi^- $ pair from $\Lambda $ decay in a $pp \pi^-$ event. The blue histogram is ${\rm ln}L$ of the incorrect pair of $p \pi^- $ in a $\Lambda $ decay. The blue histogram is vertically scaled 10 times.   }
\label{logprob}
\end{figure}

Using the distribution functions given above, the log likelihood function (${\rm ln}L$) is defined as:
\begin{equation}
  {\rm ln}L = -{\rm ln}( f_{DCA(\pi p)} \times  f_{DCA(K^- p)}  \times  f_{DCA(K^- \Lambda)}  \times  f_{DCA(\Lambda p)} \times  f_{M_{inv.\pi ^- p}} ).
\end{equation} 
${\rm ln}L$ distributions for MC simulations as well as for data are shown in Fig. \ref{logprob}.
The pairs having smaller ${\rm ln}L $ were chosen as the correct $\pi ^- p$ pair from $\Lambda$ decay; events with ${\rm ln}L > 6$ were rejected. 
Based on our simulation, the incorrect $\pi^- p$ pair selection probability is estimated to be less than 0.5 \% of the total simulated events.
Although utilization of the ${\rm ln}L $ function is valid for the $\Lambda p n$ three body state, it would be somewhat less effective for other reaction channels.

The reaction vertex was defined to be at the center of vertex($K\Lambda $) and vertex($Kp$), where vertex($K\Lambda $) is the closest approach point of the kaon and $\Lambda$ tracks on the kaon track, and vertex($Kp$) that of kaon and proton tracks on the kaon track.
The vertex was constrained to lie within the fiducial volume of the target, 30 mm in radius and 100 mm long. The contamination from reactions on the material in the target cell is estimated to be less than 2 \% of events in the fiducial volume from the empty-target data. Furthermore, by focusing on the final state $\Lambda p n$, it becomes negligible because of kinematical conservation.



\section{Results}
\begin{figure}[!hbtp]
\centering\includegraphics[width=5.5in]{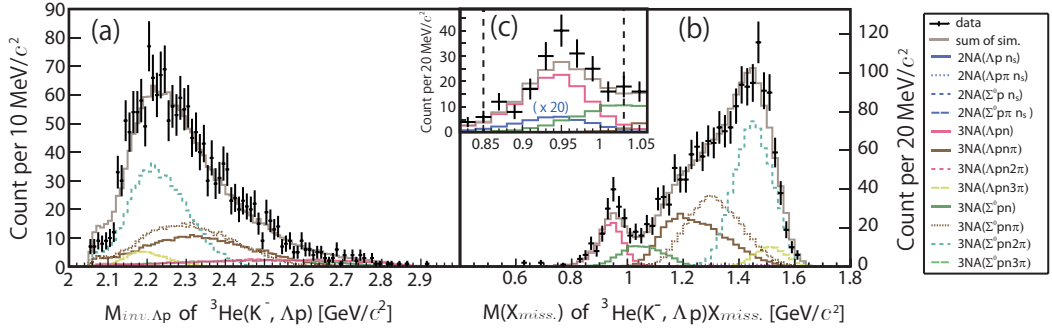}%
\caption{Inclusive spectra of the $ {\rm{}^3He}(K^-,\Lambda p)$ reaction and the global fit result of simulation with multi-nucleon absorption processes.  (a) $\Lambda p$  invariant mass distribution. (b) missing mass $M(X_{miss.})$ spectra of $ {\rm{}^3He}(K^-,\Lambda p)X_{miss.}$ and (c) the close up view of (b) around the missing neutron region. In Figure (c), 2NA$(\Lambda pn_s)$ is vertically scaled 20 times. The dashed vertical lines in (c) show $n$-window selection. }
\label{inLpall}
\end{figure}

Previous reports of an S=-1 di-baryonic state (so-called ``$\bar{K}NN$" state ) were based on  $\Lambda p $ invariant mass spectra \cite{finuda,disto}.
The present  $\Lambda p $ invariant mass spectrum, based on the above analysis, is shown in Fig. \ref{inLpall}(a). 
There are many channels having the $\Lambda p$ in the final state. 
To focus on the $\Lambda p n $ final state, the missing mass of the $ ^3{\rm He}(K^-, \Lambda p) X_{miss.}$ reaction was calculated  kinematically. A neutron peak is clearly seen as shown in Fig. \ref{inLpall}(b).
The resolution of the $\Lambda$p invariant mass $M_{inv.\Lambda p}$, and of the missing mass $M(X_{miss.})$ were
~10 MeV at $M_{inv.\Lambda p}$ $\sim$ 2.37 GeV/$c^2$ and ~40 MeV at $M(X_{miss.})$ $\sim$ 0.94 GeV/$c^2$, respectively.
To estimate the purity of our $\Lambda p n$ event selection, we applied a ``multi-channel global fit''  simultaneously to the $\Lambda p$ invariant mass and the missing mass spectra, with simulated physics processes which might contribute to these two spectra, 
namely multi-nucleon kaon absorption with
 multi-pion emission. 
We denoted each process as 2NA($YNN_s+ \#n\pi$) and 3NA($YNN+ \#n\pi$), where $Y$ is a $\Lambda $ or $\Sigma ^0$, $N$ is a nucleon, $N_s$ is a spectator nucleon, and $\#n$ is  the number of pions. For simplicity, we haven't taken into account for the final state interaction. 
In this simulation, we assumed that particles in the final state are distributed proportionally to the phase space volume in the same way as given in Eq. \ref{3NALpn}.
\begin{equation}
\label{simgene}
\frac{d^{2}\sigma_{\left( YNN +\#n\pi \right)}} {dT_{n}^{\,CM}d\cos\theta^{\,CM}_{n}}
\propto \rho_{3+\#n} \left( YNN +\#n\pi \right).
\end{equation}
 If there is a spectator nucleon, the Fermi momentum distribution is considered to reproduce the $^3$He($e,e'p$) result \cite{fermi}. 
The generated events were converted to the event data format, and examined with the same analysis routine applied for the real data.
A list of physics processes, taken into account for the global fit, is shown in  Table \ref{table_proc},
together with the relative yields obtained for each process to the number of obtained events. 
The fit results are given as histograms in Fig. \ref{inLpall}, and the two spectra are well described by the processes listed in Table \ref{table_proc}. $\chi ^2$ and DOF of the fit are 122 and 147, respectively. 
Figure \ref{inLpall}(c) shows a close up view in the missing neutron region. As shown in the figure, the $\Lambda pn$ final state events can be selected by setting the neutron window to be $0.85 < M(X_{miss.}) <1.03$ [GeV/$c^2$]. 
For simplicity, we denote those events in the neutron window as ``$\Lambda p n $ events".
This close up view also indicates that the fit result of the relative $\Lambda p n $ yield  is weaker than the data (about 83 \% compared to data). It implies that there could be an unidentified channel which contributes to the yield of the $\Lambda p n $ final state, other than 3NA($\Lambda p n $) as discussed later. From the global fit,
it is expected that three channels, 3NA($\Lambda p n $), 3NA($\Sigma^0 p n $), and 2NA($\Lambda p n_s $), will remain in the neutron window ($n$-window) at the ratio of 0.62 : 0.20  : 0.01, compared with the data in the $n$-window.

\begin{table}[!ht]
\caption{Relative yield of each component to the number of obtained events.in the global fit, normalized to that of data (same for each component in the $n$-window normalized by the yield of data in the $n$-window). 
Note that the spectral shapes of some reaction channels are quite similar, especially for the channels given in a row.}
\label{table_proc}
\centering
\begin{tabular}{|c|c|c| |c|c|c|}
\hline
process & \multicolumn{2}{c||}{relative yield} & process & \multicolumn{2}{c|}{relative yield} \\ 
\hline
 & all & $n$-window & & all & $n$-window \\     
\hline
\hline
2NA ($\Lambda pn_{s})$ & $ 0.001$  &  0.01 & 2NA ($\Sigma ^0 p n_{s}  $) & $<10^{-4}$ &  $<10^{-2}$ \\ 
\hline
2NA ($\Lambda p  n_{s} + \pi $) & $<10^{-4}$ &  $<10^{-2}$ & 2NA ($\Sigma ^0 p n_s + \pi $) & 0.010  &  $<10^{-2}$ \\ 
\hline
3NA ($\Lambda p n  $) & $0.072 $  & 0.62 & 3NA  ($\Sigma ^0 p n$) & $ 0.058 $  & 0.20\\ 
\hline
3NA ($\Lambda  p n + \pi $) & $ 0.199 $  &  $<10^{-2}$ & 3NA ($\Sigma ^0 p n + \pi $) & $ 0.239 $  &  $<10^{-2}$\\ 
\hline
3NA ($\Lambda  p n + 2\pi $) & $<10^{-4}$ &  $<10^{-2}$ & 3NA ($\Sigma ^0 p n + 2\pi $) & $ 0.354 $ &  $<10^{-2}$\\
\hline
3NA ($\Lambda  p n + 3\pi $) & $ 0.039 $ &  $<10^{-2}$ & 3NA ($\Sigma ^0 p n+ 3\pi $) & $<10^{-4}$ &  $<10^{-2}$\\
\hline
\end{tabular}
\end{table}

\begin{figure}[!hbtp]
\centering\includegraphics[width=5.0in]{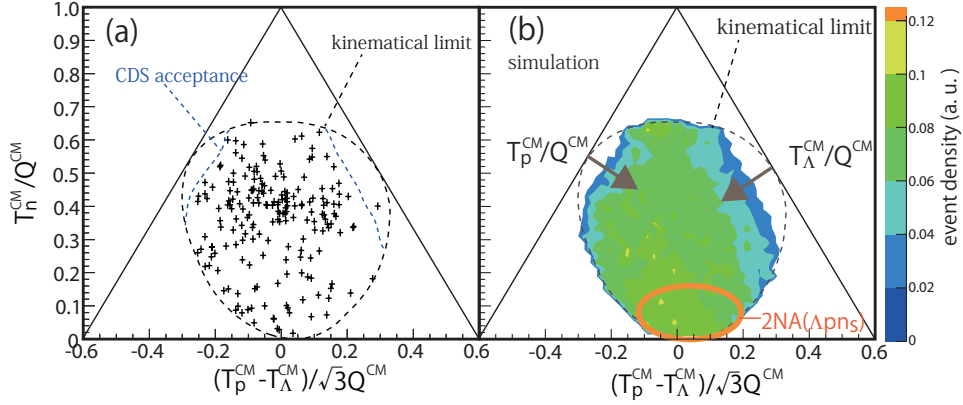}

\caption{(a)Dalitz plot of the selected $\Lambda p n$ events in the CM frame. (b) The simulated detector acceptance in a Dalitz plot.   }
\label{dalitz}
\end{figure}

The $\Lambda p n$ event distribution, over the phase space in the CM frame, is shown in 
Fig. \ref{dalitz}(a), as a Dalitz plot with kinetic energies of $\Lambda $, proton and neutron, normalized by  the $Q$-value of the reaction. 
And for reference,  the detection efficiency of the CDS is plotted in Fig. \ref{dalitz}(b).
It shows that our detector system has a fairly flat acceptance over the $\Lambda pn$ phase space, except for the upper-left and the upper-right corners. Acceptance reduction of these corners comes from a lower kinetic energy of the $p $ and $\Lambda $ (below the detection threshold). There is, however, sufficient  acceptance at the bottom, where two-nucleon kaon absorption with a spectator neutron $n_s$, $K^- + ^3$He$\rightarrow \Lambda p + n_s$, is expected.

The data events in Fig. \ref{dalitz}(a) are widely distributed over the kinematically allowed region, which is consistent with the global fit showing that the major component is coming from the three nucleon absorption process.
In the Dalitz plot, an event concentration was observed at a normalized neutron kinetic energy $T_n^{CM} / Q^{CM} \approx 0.4$, which indicates that the $\Lambda p$ invariant mass will have a structure corresponding to that energy. 

\begin{figure}[!hbtp]
\centering\includegraphics[width=5.0in]{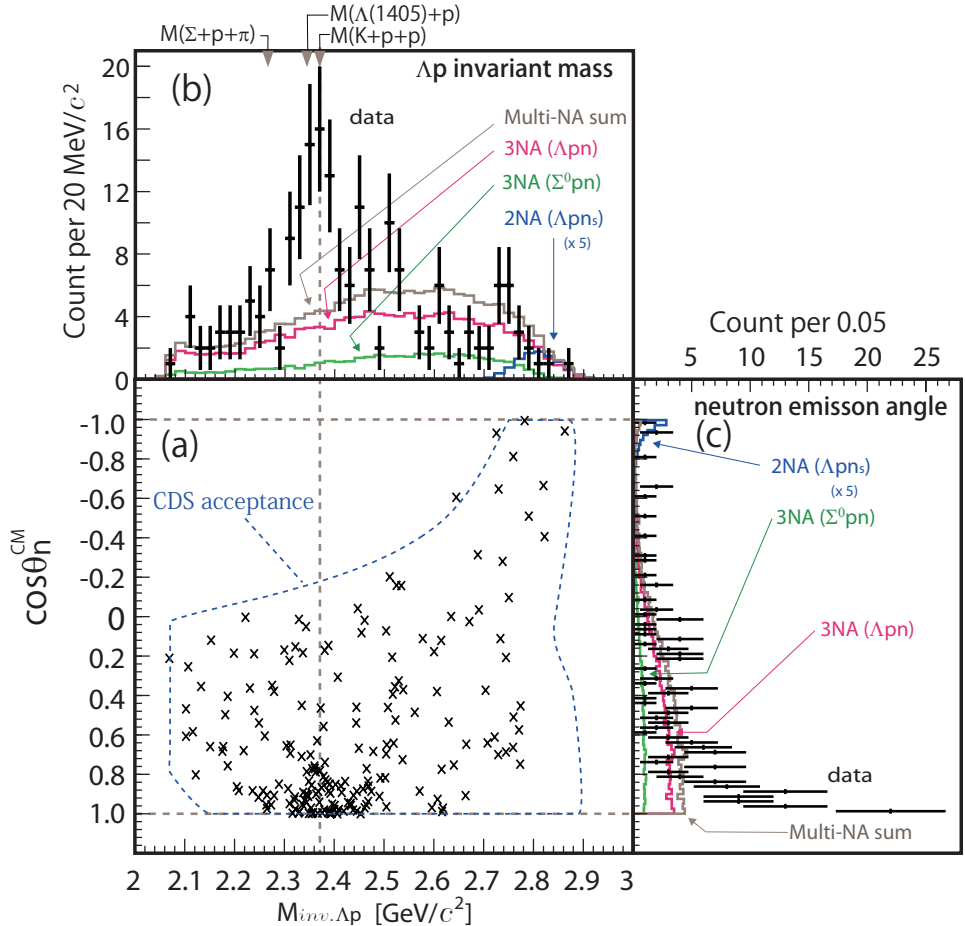}
\caption{(a) 2-D distribution of the $\Lambda p$ invariant-mass and the emission-angle of the missing neutron. The blue dashed lines show crude CDS acceptance boundary for the $\Lambda p n$ events.
(b) $\Lambda p$ invariant mass with simulated spectra obtained by the global fit in the neutron window.  (c) Angular distribution of the missing neutron, kinematically reconstructed, as a function of $\cos \theta_n^{CM}$. The histograms show the contributions of the three remaining channels in $n$-window, whose yields are given by the global fit.}
\label{datawokpp}
\end{figure}

Distribution of the $\Lambda p$ invariant mass and the calculated neutron emission-angle are shown in  Fig. \ref{datawokpp}(b) and (c), and the scatter plot of the two is given in Fig. \ref{datawokpp}(a).  
As indicated in the Dalitz plot, an unexpected peak structure is seen at $M_{\Lambda p} \sim M(K^-+p+p)$ in the $\Lambda p$ invariant mass spectrum, where the global fit is insensitive. 

As shown in Fig. 6(c), the neutrons in this peak structure are clearly concentrated at the forward region, where the momentum transfer to the $\Lambda p$ system is minimum,  
while the wide distribution can be explained by the multi-nucleon
absorption processes given by the global fit in the $n$-window.
The values obtained for $\chi ^2$ / DOF in Fig. \ref{datawokpp}(b) and (c) are 135 / 43, and 138 / 76, respectively.
This $\chi^2$ test result indicates that the observed peak structure could not be explained by multi-nucleon absorption processes.  

\section{Discussion}
To explain the excess,
the existence of a simple finite-size S-wave single-pole structure over the three-body phase space which decays to $\Lambda p$ in the final state is assumed. Thus the formation cross-section ($\sigma_X$) of the pole can be written as:
\begin{equation} \label{pole} 
 \frac{d^2 \sigma _X}{dM_{inv.\Lambda p}dq_{\Lambda p}} \propto \rho _{3}(\Lambda p n) \times \frac{(\Gamma _X/2)^2}{(M_{inv.\Lambda p}-M_{X})^2 + (\Gamma _{X} /2)^2 } \times  | \exp{(-q_{\Lambda p}^2/2Q_{X}^2)}|^2 ,
\end{equation} 
where $M_{inv.\Lambda p}$ is the invariant mass of $\Lambda p$, $q_{\Lambda p}$ is the momentum transfer of the reaction ($q_{\Lambda p}\ =\ |{\bf p}_\Lambda + {\bf p} _{p} | $),  $M_X$ is the energy, $\Gamma_X$ is the decay-width, and $Q_X$ is the form factor parameter of the pole. 
 The first term in the formula is the three-body Lorentz-invariant phase space of $\Lambda p n$, the second  the Breit-Wigner formula, and the third the square of the form-factor, which can also be interpreted as the sticking probability of a plain-wave having $q_{\Lambda p}$ to a harmonic oscillator having finite size $\approx \hbar / Q_X$.

\begin{figure}[!hbtp]
\center\includegraphics[width=5.0in]{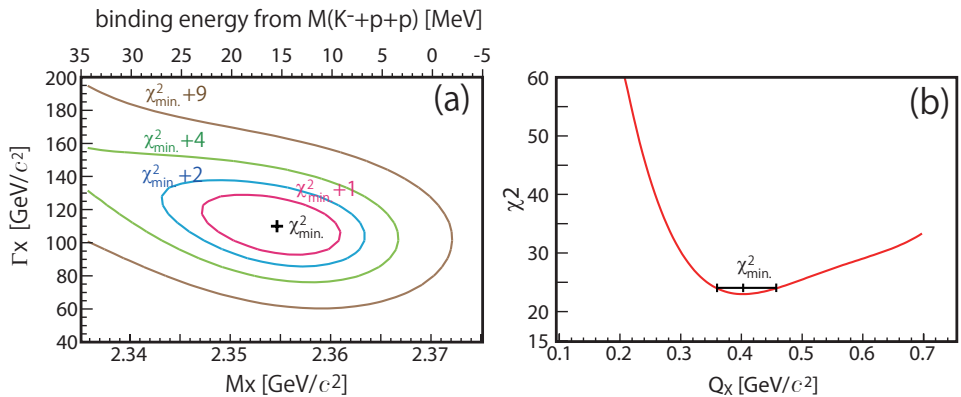}
\caption{(a) two dimensional $\chi ^2$ map of mass and width of pole structure.  (b) $\chi ^2$ distribution of $Q_X$. }
\label{fitresult1}
\end{figure}

\begin{figure}[!hbtp]
\center\includegraphics[width=5.5in]{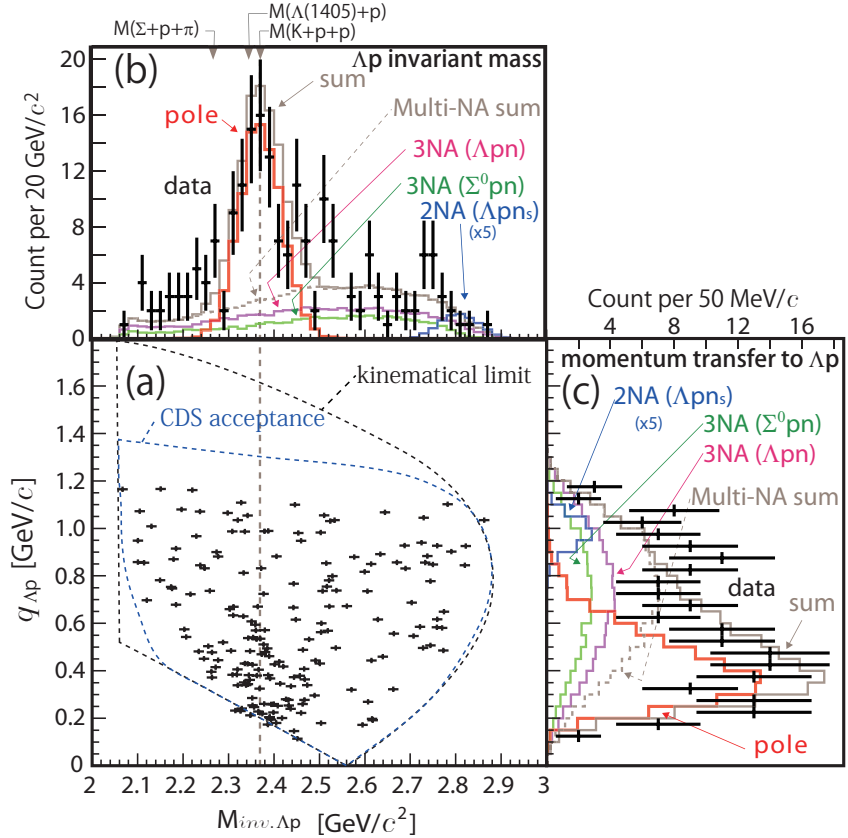}
\caption{(a) 2-D distribution of the $\Lambda p$ invariant mass and the momentum transfer to $\Lambda p$.  (b) the fit result of the $\Lambda $p invariant mass spectrum. (c) distribution of the momentum transfer to $\Lambda p$ and the fit results of simulation.}
\label{fitresult2}
\end{figure}

We generated events according to Eq. \ref{pole} in the simulation.  A $\chi^2$ comparative test was made  between the experimental data  and the simulated pole together with the multi-nucleon absorption processes. 
Thus, we fitted the spectra keeping the yield of 3NA($\Lambda p n$) as a free parameter, because  we introduced a new pole, which decays to the $\Lambda p$ final state.  The other two yields for 3NA($\Sigma^0 p n$) and 2NA($\Lambda p n_s$) are fixed as they are given by the global fit.
We first assumed  $Q_{X}= \infty $ and 
made a two-dimensional $\chi^2$ map on the $M_X$ and $\Gamma_X$ plane to define the minimum $\chi^2$ for the  invariant mass spectra.
Then, $Q_{X}$ is chosen to have minimum $\chi^2$ for the momentum transfer distribution, at a given $M_X$ and $\Gamma_X$. This process was iterated until the parameters converged. 
The two-dimensional $\chi^2$  map as a function of the $M_X$ and $\Gamma _X$ is shown in Fig.~\ref{fitresult1}(a). Figure~\ref{fitresult1}(b) plots $\chi ^2$ as a function of momentum transfer.  
The minimum $\chi ^2$ point is at $M_X =$ 2355 $ ^{+ 6} _{ - 8}$ (stat.) $\pm $ 12 (syst.)  MeV/$c^2$,  $\Gamma _X =$ 110 $ ^{+ 19} _{ - 17}$ (stat.) $\pm $ 27 (syst.) MeV/$c^2$, and $Q_X  = 400$ $^{+60}_{-40} $  (stat.) MeV/$c$. 
The statistical error is defined  as the $\chi ^2_{min.}+1$ contour. 
The systematic uncertainties are evaluated by considering the magnetic field strength in the CDS, the likelihood threshold to select the $\Lambda p$ pair, and binning of the invariant mass spectra.
The fit results are shown in Fig. \ref{fitresult2}. Because we simply assumed that the forward neutron emission is due to the form factor as it is given in Eq. \ref{pole}, we re-plotted Fig. \ref{datawokpp} in terms of $q_{\Lambda p}$ instead of $\cos \theta _n ^{CM}$. 
The values obtained for $\chi ^2$ / DOF in Fig. \ref{fitresult2}(b) and (c) are 68 / 45, and 23 / 27, respectively. 

To obtain the cross section, one needs to know the detailed event distribution and the acceptance. 
If we assume that all the angular distributions can be given by the fit results, 
the acceptance correction can be applied under this assumption. The cross sections can be evaluated as;  pole : 7 $\pm 1 $ $\mu $b, 3NA($\Lambda pn$) : 17 $\pm2 $ $\mu $b, and 2NA($\Lambda p n_s$) : 0.8 $ ^{+2.7} _{-0.8} $ $\mu $b. If we rely on the global fit and event distribution outside of the neutron window, then we can also determine the 3NA($\Sigma ^0 pn$) cross section to be 28 $\pm5 $ $\mu $b in total. The errors are the quadratic sum of the statistical and the systematic ones, where the systematic uncertainty mainly arises from the target-$^3$He and beam-kaon yields. 

There could be many interpretations for the pole found in the $\Lambda p$ invariant mass distribution in the $^3$He$(K^-, \Lambda p)n_{miss.}$ reaction channel at $p_K = 1$ GeV/$c$, even if the simplest present assumption of Eq. \ref{pole} is valid. A na\"{\i}ve interpretation of the pole would be a $\bar{K}NN$ bound state, since the pole position is located below the $M(K^- + p + p)$ threshold of 2370 MeV/$c^2$.  It could also be a shallow bound or unbound resonance of the $Y^*p$ system, because  the $Y^*p$ threshold is located at 2343 MeV/$c^2$ (assuming $Y^*=\Lambda(1405)$ at 1405 MeV/$c^2$).  However, there is no clear discrimination between the two interpretations given above, if we assume $\Lambda(1405)$ to be a $K^-p$ bound state or penta-quark like structure. 
 
The $Q_{X}$ of $\sim $ 400 MeV/$c$ is rather large compared to the Fermi-motion in light nuclei of about 100 MeV/$c$, and it implies a short interaction range of about 0.5 fm. Even if we take into account the core motion in the reaction, the effect is as small as about 20 \%\cite{harada}. 
If we assume a P-wave pole resonance instead of S-wave, then the reaction with smaller $Q_X$, namely longer interaction range,  is preferred.

It should be noted that the pole position is close to the two threshold energies, $M(K^- + p + p)$ and $M(Y^*+p)$, thus the symmetric Breit-Wigner formula could be too simple. For example, the $\bar{K}NN$-decay channel opens at the $K^- pp$ threshold, so the spectral function observed in the $\Lambda p$-decay channel could be suppressed above the corresponding thresholds.

There might be a totally different approach to account for the peak structure.  One can expect a peak-like structure by assuming $Y^* N \rightarrow  \Lambda N$ conversion, after the quasi-free $Y^*$ production by the two nucleon reaction, namely the 2NA($Y^* n p_s$) reaction followed by $Y^* p_s \rightarrow \Lambda p$ in our notation ($Y^* = \Lambda(1405)$ or $\Sigma(1385)$).  In this case, the peak position naturally depends on the momentum transfer  as $M_X(q_{\Lambda p})$ (or $\cos \theta_n^{CM}$). Although the statistics is limited, there is no clear hint of momentum dependence of the peak structure as shown both in Fig. \ref{datawokpp} and Fig. \ref{fitresult2}.


\section{Conclusion}
The  $^3$He($K^{-}$,$\Lambda p )n_{miss.}$ reaction has
been studied  with an incident $K^-$ momentum of 1.0 GeV/$c$ for the first time. 
We found that 17 $\pm2 $ $\mu $b of the $\Lambda p$ events in the neutron window can be explained by a three-nucleon absorption channel, 3NA$(\Lambda pn)$ uniformly spread over the kinamatical limit, proportional to the phase space, and without any spectator nucleon.
The existence of this channel is already quite interesting, because it seems to follow simple phase space, so that the reaction is ``point-like" as in Eq. \ref{3NALpn}.  
 In contrast, absorption by two-nucleons is rather weak compared to three-nucleons. The ratio of 2NA($\Lambda pn_s$)/ 3NA($\Lambda pn$) is found to be 0.05 or less. 


Apart from the widely distributed 3NA process, a peak structure is observed in the $\Lambda p$ invariant mass around the $\bar{K}NN$ threshold in the low momentum-transfer region.
The spectral shape has been examined assuming a single-pole existence whose final state is $\Lambda p$.
Fit results show that the pole has $M_X=$  2355 $ ^{+ 6} _{ - 8}$ (stat.) $ \pm 12$ (syst.) MeV/$c^2$, $\Gamma _X=$  110 $ ^{+ 19} _{ - 17}$ (stat.) $ \pm 27$ (syst.) MeV/$c^2$, respectively, with a Gaussian form-factor parameter $Q_X=\ 400\ ^{+60}_{-40} $MeV/$c$, having a cross section of about 7 $\mu$b.
The form factor parameter $Q_X \sim$ 400 MeV/$c$ implies that the range of interaction is about 0.5 fm.

\ack

We gratefully acknowledge all the staff members at J-PARC for their invaluable contributions.
We would like to thank Professor Toru Harada and Professor Yoshinori Akaishi for valuable comments and discussions.
This work was supported by RIKEN, KEK, RCNP,
a Grant-in-Aid for Scientific Research on Priority Areas
[No. 17070005 and No. 20028011], a Grant-in-Aid for
Specially Promoted Research [No. 20002003], a Grant-
in-Aid for Young Scientists (Start-up) [No. 20840047],
a Grant-in-Aid for Scientific Research on Innovative Areas [No. 21105003], a Grant-in-Aid for JSPS Fellows
[No. 12J10213], and the Austrian Science Fund (FWF)
[21457-N16].


%

\end{document}